\documentclass{article}
\usepackage[english]{babel}

\usepackage[letterpaper,top=2cm,bottom=2cm,left=3cm,right=3cm,marginparwidth=1.75cm]{geometry}

\usepackage{makecell}
\usepackage{booktabs}
\usepackage{comment}

\usepackage{amssymb}
\usepackage{pifont}

\usepackage{amsmath}
\usepackage{graphicx}
\usepackage[colorlinks=true, allcolors=blue]{hyperref}

\usepackage{xspace}
\usepackage{soul}

\newcommand{\etal}{\emph{et al.}\xspace}

\newcommand{\eg}{\emph{e.g.}\xspace}
\newcommand{\aka}{\emph{a.k.a.}\xspace}

\usepackage{authblk}
\author[]
       {Zhushou Tang$^{1}$\ \ Lingyi Tang$^2$\ \ Keying Tang$^{3}$\ \ Ruoying Tang$^4$ \\
       $^1$PWNZEN InfoTech Co., LTD\ \ $^2$The Webb Schools\\$^3$Shanghai Fushan Foreign Language School\ \ $^4$Shanghai Pinghe School
       }
       
\title{When Similarity Digest Meets Vector Management System: A Survey on Similarity Hash Function}

\date{}

\begin{document}
\maketitle

\begin{abstract}
The booming vector manage system calls for feasible similarity hash function as a front-end to perform similarity analysis. In this paper, we make a systematical survey on the existent well-known similarity hash functions to tease out the satisfied ones. We conclude that the similarity hash function $MinHash$ and $Nilsimsa$ can be directly marshaled into the pipeline of similarity analysis using vector manage system. After that, we make a brief and empirical discussion on the performance, drawbacks of the these functions and highlight $MinHash$, the variant of $SimHash$ and $feature\ hashing$ are the best for vector management system for large-scale similarity analysis.  
\end{abstract}

\section{Introduction}

In this section, we introduce the \emph{vector management system}, \emph{similarity hash function}, and analyze the interleaving between them.

\noindent{\bf Vector Management System.} 
The proliferation of vector management system (\eg, Milvus~\cite{wang2021milvus}, Analyticdb-v~\cite{wei2020analyticdb}, PASE~\cite{yang2020pase}, faiss~\cite{johnson2019billion}, Vearch~\cite{li2018design}, SPTAG~\cite{chen2018sptag}) satisfys the management of sheer volume of high-dimensional vectors generated by data science and AI applications.
By taking advantage of GPUs or CPUs and sorts of optimizations (\eg, indexing), these systems provide versatile similarity functions, including Euclidean, Inner product, Jaccard, Tanimoto, Hamming, SuperStructure, and SubStructure distance measurement, allowing exploring similar vectors in an effective approach. 

\noindent{\bf Similarity Hash Function.} 
The similarity hash function is wildly used in malware triage~\cite{naik2019ransomware, li2015experimental, server2015fuzzy, jang2011bitshred}, forensic analysis~\cite{seo2009detecting}, plagiarism detection~\cite{schleimer2003winnowing}. As opposed to cryptographic hash function, similarity hash function maps features of an entity (\eg, binary, text) into some high dimensional space, namely digest, since the distance between similarity hashing digests is an approximation of original features, similarity can be measured by distance within these digests. Note that Deep Supervised Hashing~\cite{luo2020survey} is not discussed here.

\noindent{\bf Gap between the Vector Management System and Similarity Hash Function.} At a first glance, the digest generated by a similarity hash function can be directly fed into vector management system for similarity analysis. However, not all digests meet the requirement of vector management system. As depicted by Digest~\ref{equ:ssdeep}, the $n$ at the beginning of the digest generated by ssdeep~\cite{kornblum2006identifying} is used for the first round comparison, comparison result is used to select one of the subsequent two non-fixed length digests for the second round comparison, such workflow increases the complexity of comparison, and the variant-length digest is not supported by the vector management system in the mean time.

\begin{equation}\tiny
\label{equ:ssdeep}
\begin{aligned}
\begin{split}
\underbrace{1536}_{\text{n}}:\underbrace{T0tUHZbAzIaFG91Y6pYaK3YKqbaCo/6Pqy45kwUnmJrrevqw+oWluBY5b32TpC0}_{\text{1st chunk}}:\underbrace{T0tU5s7ai6ptg7ZNcqMwUArKvqfZlMC0}_{\text{2nd chunk}}
\end{split}
\end{aligned}
\end{equation}

\noindent{\bf Overlap of the Vector Management System and Similarity Hash Function.}

It is observed that there are overlapping functions between the vector management system and similarity hash function. For example, the \emph{Locality-Sensitive Hashing} (LSH) is used by both some similarity hash functions (\eg, SimHash) and vector management system (\eg, indexing of faiss), the overlapping approximation makes the analysis result unpredictable.

The main goal of this work is to make a systematical survey on the existent similarity hash functions and figure out these can be used as a front-end of the vector management system. 

\section{Survey on the Similarity Hash Functions}
In a word, the ideal digest generated by a similarity hash function should be fixed-length such that can be fed into the vector management system and the distance measurement function should fall into one of the functions supported by the vector management system such can take advantage of these systems.

This section starts with building new triage for these functions, then dives into each similarity hash function and briefly summarizes the functions.

\subsection{New triage}

In general, the similarity hash function consists two parts: 
{\it (i)} digest generation and 
{\it (ii)} digest comparison. 
The digest generation is further separated into feature extraction and feature encoding process. 
Based on how the digest function encode the features and compute the final digest,
the current triage of digest function can be separated into: \emph{Context-Triggered Piecewise Hashing} (CTPH, namely fuzzy hash), \emph{Block-Based Hashing} (BBH), \emph{Locality-Sensitive Hashing} (LSH),  \emph{Statistically-Improbable Features} (SIF). 
For example, Oliver~\etal~\cite{oliver2013tlsh} categorized the similarity digest function into three families: CTPH, BBH, and SIF; 
Li~\etal~\cite{li2015experimental} separate them into CTPH and BBH.
Besides, Gayoso~\etal~\cite{gayoso2014state} add an additional item, namely~\emph{Block-Based Rebuilding} (BBR), to the categories. 

Other than the aforementioned categories, we build new triage based on how the digest compared.  
Considering that the vector management system only accepts/operates fixed-length bit-vector/floating-vector and complex similarity analysis is not supported yet, we separate these functions into three families: 
{\it (i)} \emph{Vector-Hashing} (VH), 
{\it (ii)} \emph{Portable-Vector-Hashing} (PVH) and 
{\it (iii)} \emph{Non-Vector-Hashing} (NVH). 
The VH family is that the vector can be fed into vector management system directly and exploit the similarity function within the vector management system; 
PVH family digest contains auxiliary information, but can be tailored to fit the vector management system (driven by extra control);  
At last, the non-fixed length digest or complex similarity analysis which is not supported by vector management system is categorized into NVH family.
For example, the ssdeep~\cite{kornblum2006identifying} is categorized as CTPH in conventional, however we classify it as NVH, for the comparison of this function uses \emph{weighted edit distance} between two digests which is not supported by the vector management system.

\subsection{Similarity Hash Function Analysis}

\begin{table}[h]\scriptsize
\centering
\begin{tabular}{ 
m{0.9cm}<{\centering} m{0.9cm}<{\centering} m{2.0cm}<{\centering} m{2.4cm}<{\centering} m{2.0cm}<{\centering} m{2.0cm}<{\centering} m{1.1cm}<{\centering} m{0.6cm}<{\centering} 
}
Scheme & Original Family & Feature Extraction & Feature Generation & Digest Property & Distance Function & New Family & Designed For \\\toprule
MinHash \cite{broder1997resemblance} & LSH & Winnowing & Bucket mapping & - & Jaccard Similarity & VH & Binary \\\hline
dcfldd \cite{nicholas2002dcfldd} & BBH & Split the data into sectors or blocks of fixed-length. & Compute the corresponding cryptographic hash value for each of these blocks. & - & - & NVH & Binary \\\hline
Nilsimsa \cite{damiani2004open} & LSH & Winnowing & Bucket mapping (Pearson hash~\cite{pearson1990fast}) followed by encoding the accumulated value to bit-vector under the control of a threshold. & 32-bytes length vector & Hamming distance & VH & Text \\\hline 
ssdeep \cite{kornblum2006identifying} & CTPH & Use Alder-32 (rolling hash) to identify boundaries of a chunk. & Use FNV-hash~\cite{fowler1991fowler} upon the chunk and use the last 6-bits. & Non-fixed length with auxiliary information, ASCII encoding & Weighted edit distance after the auxiliary information comparison & NVH (variant length digest) & Text \\\hline 
SimHash \cite{sadowski2007simhash} & LSH & Accumulate the occurrence of pre-defined 16 8-bits tags (Shingling). & The combination of sum table & 32-bits vector enclosing auxiliary information (\eg, file extension). & Hamming distance & PVH & Text \\\hline 
sdhash \cite{roussev2010data} & SIF & Get entropy estimates of each 64-bytes block, use winnowing to select the entropy estimates. & Bloom filter (SHA-1) & Multiple bloom filters & Average the maximums of per-filter in one set `AND' each counterpart filters. & NVH & Text \\\hline 
TLSH \cite{oliver2013tlsh} & LSH & Winnowing & Bucket mapping (Pearson hash~\cite{pearson1990fast}) followed by encoding the accumulated value to bit vector under control of quartile points. & checksum + length + quartile points + encoded value & Circular distance + auxiliary information & VNH & Binary \\\hline
mvHash-B \cite{breitinger2013mvhash} & BBH & Majority vote followed by RLE (length encoding algorithm) & Bloom filter & Multiple bloom filters & Average the minimums of per-filter in one set `XOR' each counterpart filters. & NVH & Binary \\


\bottomrule

\end{tabular}
\caption{\label{tab:scheme}Break down of similarity hash functions.}
\end{table}

To categorize the similarity hash functions, we go through the well-known digest generation schemes, roughly summarize each similarity hash function, and tease out the functions that can be marshaled into the pipeline of similarity analysis using vector management system. The schemes are listed in Table~\ref{tab:scheme}. The original family the function belongs to is listed in the second column of the table, break downs of the scheme (\eg, feature extraction, feature generation, digest style, and distance measurement function) are depicted in 3-6 columns. The seventh column is the new family assigned for each scheme. 
Note that, although there are multiple variants for a given scheme, we only discuss the well-known version (\eg, the SimHash~\cite{simhash_variant} proposes to use Shingling as a prerequisite for feature extraction, the open source implementation on github uses Jieba instead; Although FKSum~\cite{chen2008efficient}, SimFD~\cite{seo2009detecting}, MRSH~\cite{roussev2007multi, breitinger2012similarity}) improve either the efficiency or precision of ssdeep, we take in ssdeep nonetheless). Details of these schemes are also summarized by Gayoso \etal ~\cite{gayoso2014state}, we encourage readers to refer this material.





\section{Discussion}

In this section, we discuss the feasible similar hash function, we conclude that industrial adopts to use high efficient function and domain knowledge plays an import role in similarity analysis. 

\subsection{The Vector-Hashing and Portable-Vector-Hashing}

From table~\ref{tab:scheme}, we find that MinHash and Nilsimsa use bucket mapping to align the digest to fixed-length, such that can be marshaled into the pipeline of similarity analysis using vector manage system. By excluding the auxiliary information, SimHash can be ported to the pipeline too. We also notice that TLSH is the variant Nilsimsa, the auxiliary information appended to digest improves the precision, but does not fit the vector manage system. 

\subsection{The Non-Vector-Hashing}
The distance measurement between two bloom filters set is conducted by a filter-to-filter comparison with complexity $O(m*n)$, the comparison is not supported by vector manage system, such that sdhash and mvHash-B can not be plugged into  the pipeline. ssdeep uses non-fixed length digests for the second round comparison, also leading to failure of similarity analysis.

\subsection{Tradeoff Between Precision and Performance}

Lots of work has evaluated the precision of the similar hash functions~\cite{charikar2002similarity, breitinger2012security, oliver2014using, kim2020revisiting}, however, the sheer volume of high-dimensional vectors generated by data science and AI applications allow for a certain tolerance on precision. For example, even though the ssdeep is the de facto standard of similarity hash function and integrated into VirusTotal, there is no sign indicate VirusTotal has exploited the digests to find similar binaries. On the other side, the lightweight Simhash is announced to be used by Google for duplicate detection for web crawling~\cite{manku2007detecting}, Minhash is used for Google News personalization~\cite{das2007google}, Jang \etal~\cite{jang2011bitshred} also stat that $feature\ hashing$ which maps each feature into exactly one bit in the bit array works well in large-scale malware similarity analysis. Such lightweight schemes bring extra bonus for analysis task, for example,  the $feature\ hashing$ solution can pinpoint the commonality and difference between malware samples in the mean time.

\subsection{Role of Domain Knowledge}

Although there are literals argue the universality of similarity hash functions~\cite{gayoso2014state}, we believe domain knowledge (\aka semantic) plays an important role in similarity analysis. Taking android third-party library correlation as an example, different  arguments of compilation will dramatically change the binary, leading to similarity detection failure, but works~\cite{ma2016libradar, tang2019securing, feichtner2019obfuscation} take in domain knowledge (\eg, System Call, Abstract Syntax Tree) as feature outperform those without domain knowledge.  

\bibliographystyle{alpha}
\bibliography{main}

\end{document}